\documentclass{article}
\usepackage{FPSpro}
\begin{document}
\title{ THE HIGHEST ENERGY PARTICLES IN THE UNIVERSE: THE MYSTERY
AND ITS POSSIBLE SOLUTIONS
  }
\author{
  Pasquale Blasi        \\
  {\em Osservatorio Astrofisico di Arcetri, Largo E. Fermi, 5 - 
Firenze (ITALY)} 
}
\maketitle
\baselineskip=11.6pt
\begin{abstract}
The universe should be dark at energies exceeding $\sim 5\times 10^{19}$ eV.
This simple but solid prediction of our best known particle physics is not 
confirmed by observations, that seem to suggest a quite different picture. 
Numerous events have in fact been detected in this energy region, with 
spectra and anisotropy features that defy many conventional and unconventional
explanations. 
Is there a problem with known physics or is this a result of astrophysical
uncertainties? Here we try to answer these questions, in the light of present
observations, while discussing which information future observations may 
provide on this puzzling issue.
\end{abstract}
\baselineskip=14pt
\section{Introduction}
One of the major goals of cosmic ray physics has always been the 
discovery and understanding of the {\it end of the cosmic ray spectrum}.
Until the end of the '60s, this search was mainly aimed to understand
the limits to the acceleration processes and nature of the sources
responsible for the production of the particles with the highest energies. 
However, after the discovery of the cosmic microwave background (CMB), it 
became soon clear that the observed spectrum of the cosmic radiation had 
to be cut off at a ``natural'' energy, even if an ideal class of sources 
existed, able to accelerate particles to infinite energy. In fact, if the 
sources are distributed homogenously in the universe, the photopion 
production in the scattering of particles off the CMB photons imply a cutoff 
in the observed spectrum of the cosmic rays at an energy 
$E_{GZK}\sim (4-5)\times 10^{19}$ eV, close to the kinematic threshold 
for that process \cite{greisenzk66}. This cutoff has become known as the 
{\it GZK cutoff} and particles with $E>E_{GZK}$ are usually named ultra-high
energy cosmic rays (UHECRs).

Several experiments have been operating to detect the flux of UHECRs, starting
with Volcano Ranch \cite{linsley} and continuing with Haverah Park 
\cite{watson91} and Yakutsk \cite{efimov91} to the more recent 
experiments like AGASA \cite{agasa01,tak99,tak98,ha94}, Fly's Eye 
\cite{bird93,bird94,bird95} and HiRes \cite{kieda99}.
The search for the GZK cutoff, instead of confirming the simple picture
illustrated above, has provided stronger and stronger evidence for the 
existence of events corresponding to energies well in excess of $E_{GZK}$:
the GZK cutoff has not been found.

This problem hides many issues on plasma
physics, particle physics and astrophysics, that in their whole
represent the puzzle of UHECRs. In section \ref{sec:observations} the 
present status of observations of UHECRs is summarized; 
in section \ref{sec:serious} some caveats in the arguments that are 
often used to address the problem of UHECRs are considered, for the 
purpose of stating the problem in a clear way.
In sections \ref{sec:acceleration} and \ref{sec:td} 
the bottom-up and top-down models of UHECR origin are summarized.
In section \ref{sec:liv} some speculations are discussed 
of new physics scenarios that might play a role not only for the explanation 
of UHECRs, but also for the understanding of other current puzzles in 
high energy astrophysics. Conclusions are reported in section 
\ref{sec:conclusion}.

\section{Observations}\label{sec:observations}

The cosmic ray spectrum is measured from fractions of GeV to a (current)
maximum energy of $3\times 10^{20}$ eV. The spectrum above a few GeV and up to 
$\sim 10^{15}$ eV (the knee) is measured to be a power law with slope
$\sim 2.7$, while at higher energies and up to $\sim 10^{19}$ eV 
(the ankle) the spectrum has a steeper slope, of $\sim 3.1$. At energy 
larger than $10^{19}$ eV a flattening seems to be present. 

The statistics of events is changing continuously: the latest analysis of
the ``all experiments'' statistics was carried out in \cite{uchihori00}
where 92 events were found above $4\times 10^{19}$ eV. 47 events were 
detected by the AGASA experiment. A more recent analysis \cite{agasa01} 
of the AGASA data, carried out expanding the acceptance angle to $\sim 60^o$, 
has increased the number of events in this energy region to 59.

In \cite{tak99} the directions of arrival of the AGASA events 
(with zenith angle smaller than $45^o$) above 
$4\times 10^{19}$ eV were studied in detail: no appreciable departure 
from isotropy was found, with the exception of a few small scale 
anisotropies in the form of doublets and triplets of events within 
an angular scale comparable with the angular resolution of the experiments 
($\sim 2.5^o$ for AGASA). This analysis was repeated in \cite{uchihori00}
for the whole sample of events above $4\times 10^{19}$ eV, and a total of 12 
doublets and 3 triplets were found within $\sim 3^o$ angular scales. 
The attempt to associate these multiplets with different types of local 
astrophysical sources possibly clustered in the local supercluster did 
not provide evidence in that direction \cite{stanev}. 

Recently, the AGASA collaboration reported on the study of the small scale 
anisotropies in the extended sample of events with zenith angle $<60^o$: 
5 doublets (chance probability $\sim 0.1\%$) and 1 triplet (chance
probability $\sim 1\%$) were found. 

The information available on the composition of cosmic rays at the highest 
energies is quite poor. A study of the shower development was possible only 
for the Fly's Eye event \cite{bird95} and disfavors a photon primary 
\cite{halzen}. A reliable analysis of the composition is however 
possible only on statistical basis, because of the large fluctuations 
in the shower development at fixed type of primary particle.

The Fly's Eye collaboration reports of a predominantly heavy composition
at $3\times 10^{17}$ eV, with a smooth transition to light composition
at $\sim 10^{19}$ eV. This trend was later not confirmed by AGASA 
\cite{ha94,yoda98}.

Recently in ref. \cite{zas2000} the data of the Haverah Park experiment 
on highly inclined events were re-analyzed: this new analysis results in
no more than $30\%$ of the events with energy above $10^{19}$ eV being 
consistent with photons or iron (at $95\%$ confidence level) and no
more than $55\%$ of events being photons above $4\times 10^{19}$ eV.

Recently a new mass of data has been presented by the HiRes experiment
\cite{hiresICRC}. Only two events with energy above $10^{20}$ eV have been 
detected by this experiment insofar, compatible with the presence of a GZK 
cutoff. This discrepancy with the results of several years of AGASA operation
needs further investigation. Several 
sistematics have been identified that might considerably affect the 
determination of the energies and fluxes of fluorescence experiments 
versus the ground array techniques \cite{agasa01}. These issues 
will not be discussed further in the present paper.

\section{The GZK cutoff: how serious is its absence?}\label{sec:serious}

The puzzle of UHECRs can be summarized in the following points:

\begin{itemize}
\item {\it \underline{The production problem}:}
the generation of particles of energy $\geq 10^{20}$ eV requires
an excellent accelerator, or some new piece of physics that allows the
production of these particles in a non-acceleration scenario.

\item {\it \underline{The large scale isotropy}:}
observations show a remarkable large scale isotropy of the arrival 
directions of UHECRs, with no correlation with local structures
(e.g. galactic disk, local supercluster, local group).

\item {\it \underline{The small scale anisotropy}:}
the small (degree) scale anisotropies, if confirmed by further upcoming 
experiments, would represent an extremely strong constraint on the type of
sources of UHECRs and on magnetic fields in the propagation volume.

\item {\it \underline{The GZK feature}:}
the GZK cutoff is mainly a geometrical effect: the number of sources
within a distance that equals the pathlength for photopion production is 
far less than the sources that contribute lower energy particles, having 
much larger pathlength (comparable with the size of the universe).
The crucial point is that the cutoff is present even if plausible nearby UHECR
engines are identified.

\item {\it \underline{The composition}:}
it is crucial to determine the type of particles that generate the events 
at ultra-high energies. The composition can be really considered a smoking gun
either in favor or against whole classes of models.
\end{itemize}

The five points listed above are most likely an oversimplification of the 
problem: some other issues could be added to the list, such as the spectrum, 
but at least at present this cannot be considered as a severe constraint. 
On the other hand, specific models make specific predictions on the spectral 
shape, so that when the results of future observations will be available, this 
information will allow a strong discrimination among different explanations 
for the origin of UHECRs. Any model that aims to the explanation of the 
problem of UHECRs must address all of the issues listed above (and possibly
others).

In this section we consider in some more detail the issue of the GZK
cutoff and the seriousness of its absence in the observed data. 

It is often believed that the identification of one or a class of nearby
UHECR sources would explain
the observations and in particular the absence of the GZK cutoff.
This is not necessarily true. The (inverse of the) lifetime of 
a proton with energy $E$ is plotted in fig. 1 (left panel) together with
the derivative with respect to energy of the rate of energy losses $b(E)$
(right panel) [the figure has been taken from ref. \cite{bereICRC}]. 
The flux per unit solid angle at energy $E$ in some direction is proportional 
to $n_0 \lambda(E) \Phi(E)$, where $n_0$ is the density of sources (assumed 
constant), $\lambda(E)=c/((1/E)dE/dt)$ and $\Phi(E)$ is the source spectrum.
This rough estimate suggests that the ratio of detected fluxes (multiplied
as usual by $E^3$), at energies $E_1$ and $E_2$ is 
\begin{equation}
{\cal R}=
\frac{E_1^3 F(E_1)}{E_2^3 F(E_2)}\sim \frac{\lambda(E_1) \Phi(E_1) E_1^3}
{\lambda(E_2)\Phi(E_2) E_2^3} =
\frac{\lambda(E_1)}
{\lambda(E_2)} \left(\frac{E_1}{E_2}\right)^{3-\gamma},
\label{eq:ratio}
\end{equation}
where in the last term we assumed that the source spectrum is a power law
$\Phi(E)\sim E^{-\gamma}$.
If for instance one takes $E_1=10^{19}$ eV (below $E_{GZK}$) and 
$E_2=3\times 10^{20}$ eV (above $E_{GZK}$), 
from fig. 1 one obtains that ${\cal R}\sim 80$ for $\gamma=3$ and 
${\cal R}\sim 10$ for $\gamma=2.4$. The ratio ${\cal R}$ gives a rough
estimate of the suppression factor at the GZK cutoff and its dependence
on the spectrum of the source. For flat spectra ($\gamma\leq 2$) the 
cutoff is less significant, but it is 
more difficult to fit the low energy data \cite{blanton} 
(at $E\sim 10^{19}$ eV). Steeper
spectra make the GZK cutoff more evident, although they allow an easier 
fit of the low energy data.
The simple argument illustrated above can also be interpreted in an alternative
way: if there is a local overdensity of sources by a factor $\sim {\cal R}$,
the GZK cutoff is attenuated with respect to the case of homogeneous 
distribution of the sources. 
\begin{figure}[t]
  \vspace{7.0cm}
  \includegraphics{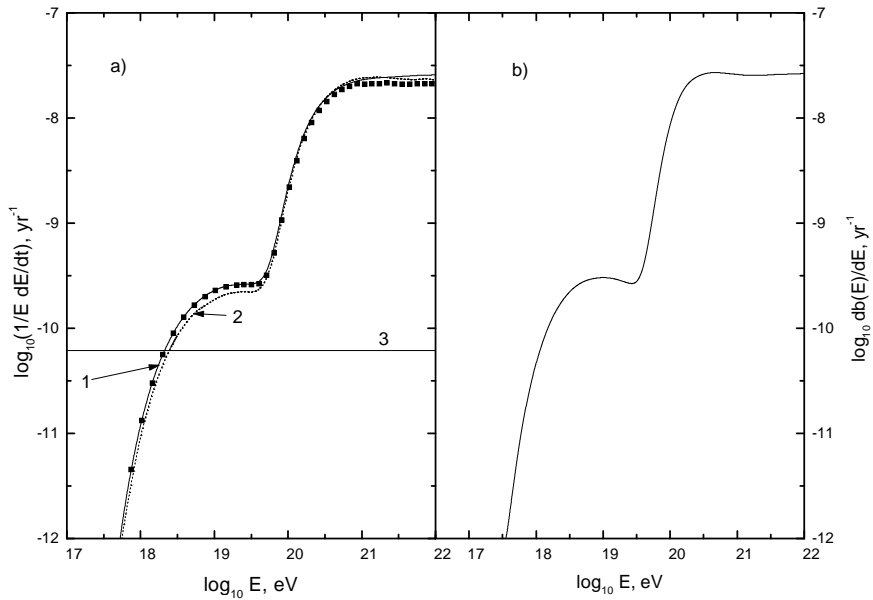}
  \caption{\it
    From \cite{bereICRC}. Left panel) $(1/E)dE/dt$ for a proton in 
\cite{bereICRC} (curve 1), in \cite{grigo88} (curve 2) and in \cite{stanev00}
(black squares). The curve 3 is the contribution of the red shift. Right panel)
The derivative $db(E)/dE$, with $b(E)=dE/dt$ at $z=0$.
    \label{fig1} }
\end{figure}
The question of whether we are located in such a large overdensity of 
sources was recently addressed, together with the propagation of UHECRs,
in \cite{blanton}. Assuming that the density of the (unknown) sources
follows the density of galaxies in large scale structure 
surveys like PSCz \cite{pscz} and Cfa2 \cite{cfa2}, the authors estimate
the local overdensity on scales of several Mpc to be of order $\sim 2$,
too small to compensate for the energy losses of particles with energy
above the threshold for photopion production. 

There is however another issue that the calculations in \cite{blanton}
address, which is related to the statistical fluctuations induced by
the process of photopion production. The large inelasticity of this process
can be taken into account properly only through the use of Montecarlo 
calculations.
When the Montecarlo is applied to simulate the small statistics of
events typical of current experiments, the fluctuations in the simulated
fluxes above $\sim 10^{20}$ eV are very large, so that for flat spectra
($\sim E^{-2}$) the discrepancy between observations and simulations on 
the total number of events above $\sim 10^{20}$ eV is at the level of 
$\sim 2 \sigma$ (in agreement with the conclusions of Ref. 
\cite{bac_wax}). The situation is represented in fig. 2
\cite{blanton} where the hatched regions show the uncertainties in the 
simulated fluxes. The data points are from AGASA \cite{tak98,hayashida99}. 
\begin{figure}[t]
  \vspace{7.0cm}
  \includegraphics{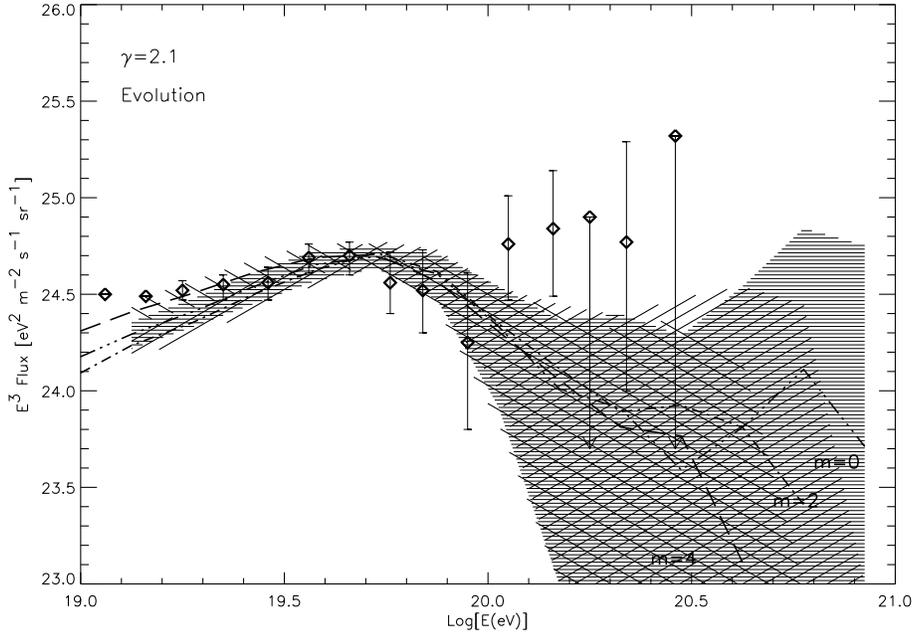}
  \caption{\it Simulated fluxes with the AGASA statistics and $\gamma=2.1$.
The sources are homogeneously distributed up to a maximum redshift 
$z_{max}=1$. The different hatches refer to different cases of evolution 
of the sources \cite{blanton}.
    \label{fig2} }
\end{figure}
The bottom line of this section can be summarized in the following few points:

1) the GZK cutoff is not avoided by finding sources of UHECRs that lie 
within the pathlength of photopion production, unless these sources are
located only or predominantly nearby and are less abundant at large 
distances. 

2) The significance of the GZK feature depends on the fluctuations in the 
photopion production, and can be addressed properly only with a enhanced
statistics of events with energy $\geq 10^{20}$ eV.

\section{The UHECRs engines}

Models for the origin of UHECRs can be strongly constrained 
on the basis on the criteria illustrated in the previous section.
The challenge to conventional acceleration models, that are supposed to work 
at lower energy scales, induced an increasing interest for more exotic 
generation mechanisms, eventually requiring new particle physics. 
In this section the main ideas on production scenarios and their signatures
will be summarized.

\subsection{\underline{Acceleration scenarios: the physics of Zevatrons}}
\label{sec:acceleration}

The accelerators able to reach maximum energies of order 1 ZeV have been
named Zevatrons \cite{olinto}. The challenge for Zevatrons was recently 
discussed in detail by several authors \cite{blandford,olinto}: the main 
concept in this class of models is that the energy flux embedded in a 
macroscopic motion or in magnetic fields is partly converted into energy 
of a few very high energy particles. This is what happens for instance in 
shock acceleration. 

A discussion of all the models in the literature is not the purpose of the
present paper, and in a sense we think it may not be very interesting. 
Nevertheless, it is instructive to understand at least which classes of
models may have a chance to explain the acceleration to ZeV energies. In
this respect, a pictorial way of proceeding is based on what is known as
the Hillas plot \cite{hillas}. Our version of it is reported in fig. 4.

\begin{figure}[t]
  \vspace{7.0cm}
  \includegraphics{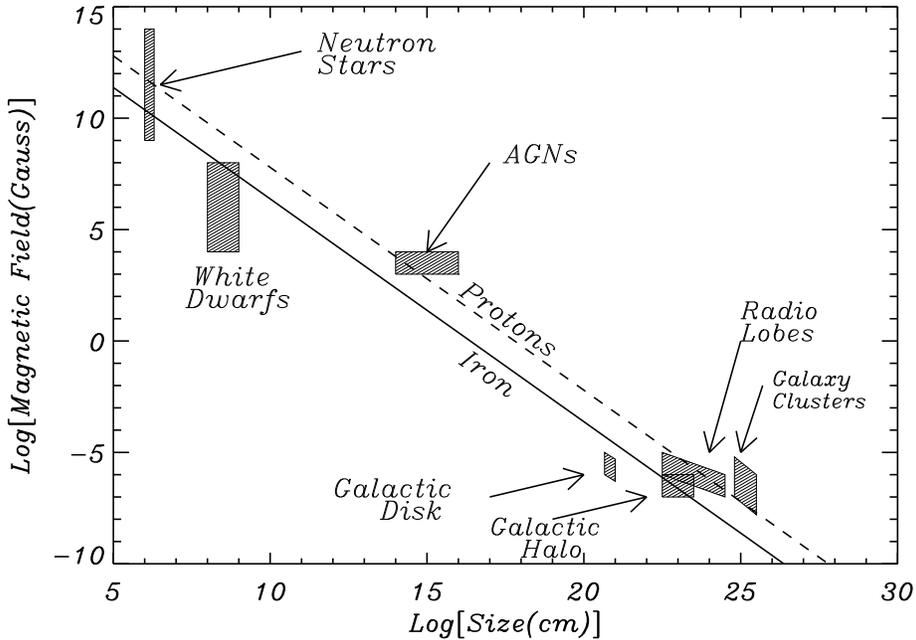}
  \caption{\it The Hillas Plot.
    \label{fig4} }
\end{figure}

In the Hillas plot the maximum energy is taken to be in its simplest form,
as determined by the local magnetic field $B$ and the size $L$ of the
accelerator: $E_{max}=Ze B L$. Here $Ze$ is the electric charge of the 
accelerated particles. From fig. 4 it is evident that only 4 classes of
sources have the potential to accelerate protons to ultra-high energies:
1) Neutron stars; 2) Radio Lobes; 3) Active Galactic Nuclei; 4) Clusters
of Galaxies. In the case of Iron, the situation becomes more promising for 
other sources, like the galactic halo or extreme white dwarfs. These sources 
would however have other problems that make them unlikely sources of 
UHECRs.

The hillas plot does not include the effect of energy losses in the 
acceleration sites. Photopion production limits the maximum energy 
achievable in clusters of galaxies to $\leq \rm{a~few}~10^{19}$ eV. 
These sources will therefore not be considered any longer as sources of 
cosmic rays above the GZK energy.
We also do not discuss here the so-called bursting sources. The prototypical
example of these sources are gamma ray bursts, that have been proposed
as sources of UHECRs \cite{GRB}. 
We refer the reader to recent literature treating this topic 
\cite{dar}.

In the following we briefly summarize the situation with the other three
classes of objects listed above. 

\underline{{\it Neutron Stars}}

The possibility that neutron stars may be accelerators of UHECRs was
discussed in detail in Ref. \cite{bible} (and references therein). The
main problem encountered in reaching the highest energies is related
to the severe energy losses experienced by the particles in the acceleration
site \cite{venkatesan,bible}. Most of the mechanisms discussed in the 
literature refer to acceleration processes in the magnetosphere of the
neutron star, where curvature radiation limits the maximum energy to 
a value much smaller that $10^{20}$ eV. 

An alternative approach is to think of acceleration processes that occur
outside the light cylinder of young neutron stars \cite{boe,lazarian}. 

Rapidly rotating, newly formed neutron stars can induce the acceleration 
of iron nuclei through MHD winds outside the light cylinder \cite{boe}.
Although the mechanism through which the rotation energy of the star is
converted into kinetic energy of the wind is not yet completely understood,
it seems from the observations of the Crab nebula that a relativistic wind
does indeed exist, with a Lorentz factor of $\sim 10^7$ \cite{begelman}.
Possible nuclei with charge $Z_{26}=Z/26$ can be accelerated in young
neutron stars to a maximum energy $E_{max}=8\times 10^{20} Z_{26} B_{13} 
\Omega_{3k}^2$, as estimated in Ref. \cite{boe}. Here $B_{13}$
is the surface magnetic field in units of $10^{13}$ G and $\Omega_{3k}=
\Omega/3000\rm{s}^{-1}$ is the rotation frequency of the star. Energies
gradually smaller are produced while the star is spinning down, so that a 
spectrum $\sim E^{-1}$ is produced by a neutron star. 
The process of escape of the accelerated particles becomes
efficient about a year after the neutron star birth. Particles that are 
generated earlier cannot escape, but can produce high energy neutrinos in 
collisions with the ambient particles and photons \cite{bedn}.
The issue of the anisotropy due to the galactic disk is currently under 
investigation \cite{aniso}.

\underline{{\it Active Galactic Nuclei}}

Active galaxies are thought to be powered by the accretion of gas onto
supermassive black holes. Acceleration of particles can occur in standing
shocks in the infalling gas or by unipolar induction in 
the rotating magnetized accretion disk \cite{thorne}. In the former scenario
energy losses and size of the acceleration region are likely to limit the
maximum energy of the accelerated particles to $\ll 10^{20}$ eV. 
In the latter case, the main limiting factor in reaching the highest
energies is represented by curvature energy losses, that are particularly
severe \cite{bible} unless moderately high magnetic fields can be kept with 
a small accretion rate. This may be the case of dormant supermassive 
black holes, possibly related to the so-called dark massive objects (DMO). 
Some 32 of these objects have been identified in a recent survey 
\cite{magorrian} and 14 of them have been estimated to have
the right features for acceleration of UHECRs \cite{boldt}. Had this model
to be right, it would not be suprising that bright counterparts to the 
UHECR events were not found, since DMOs are in a quiescent stage of their 
evolution.

Very little is known of DMOs as cosmic ray accelerators: the spectrum is 
not known, and neither is known their spatial distribution. It is therefore
hard to say at present whether DMOs can satisfy the criteria listed
in section \ref{sec:serious}. In \cite{levinson} an interesting 
prediction was proposed: if UHECRs are accelerated by unipolar induction, they
have to radiate part of their energy by synchrotron emission, resulting 
in the sources to become observable at TeV energies. 

\underline{{\it Jets and lobes}}

One of the most powerful sites for the acceleration of UHECRs is the 
termination shock of gigantic lobes in radio galaxies. Of particular
interest are a subclass of these objects known as Fanaroff-Riley class II
objects (FR-II), that can in principle accelerate protons to $\sim 10^{20}-
10^{21}$ eV and explain the spectrum of UHECRs up to the GZK cutoff 
\cite{biermann}. These objects are on average on cosmological distances.
The accidental presence of a nearby source of FR-II type might explain the
spectral shape above the GZK energy, but it would not be compatible with 
the observed anisotropy \cite{slb,blasiolinto99}. Nevertheless, it has been
recently proposed that a nearby source in the Virgo cluster (for instance M87)
and a suitable configuration of a magnetized wind around our own Galaxy might
explain the spectrum and anisotropy at energies above $\sim 10^{20}$ eV
\cite{ahn} as measured by AGASA. This conclusion depends quite sensibly 
on the choice of the geometry of the magnetic field in the wind. Several
additional tests to confirm or disprove this model need to be carried out.

\subsection{\underline{The Top-Down Approach}}
\label{sec:td}

An alternative to acceleration scenarios is to generate UHECRs by the 
decay of very massive particles. 
In these {\it particle physics inspired models}
the problem of reaching the maximum energies is solved {\it by construction}. 
The spectra of the particles generated in the decay are typically 
flatter than the astrophysical ones and their composition at the 
production point is dominated by gamma rays, although propagation effects 
can change the ratio of gamma rays to protons. The gamma rays generated 
at distances larger than the absorption length produce a cascade at low 
energies (MeV-GeV) which represents a powerful tool to contrain TD models 
\cite{bs00}. There are basically two ways of generating the very 
massive particles and make them decay at the present time: 
1) trapping them inside topological 
defects; 2) making them quasi-stable (lifetime larger than the present age 
of the universe) in the early universe. 
We discuss these two possibilities separately in the next two sections.

\subsubsection{Topological Defects}

Symmetry breakings at particle physics level are responsible for the
formation of cosmic topological defects (for a review see Ref. 
\cite{vilshe94}). Topological defects as sources of UHECRs 
were first proposed in
the pioneering work of Hill, Schramm and Walker \cite{hsw87}. The general idea 
is that the stability of the defect can be locally broken by different types of
processes (see below): this results in the false vacuum, trapped within the 
defect, to fall into the true vacuum, so that the gauge
bosons of the field trapped in the defect acquire a mass $m_X$ and decay. 

Several topological defects have been studied in the literature: ordinary
strings \cite{rana90}, superconducting strings \cite{hsw87},
bound states of magnetic monopoles \cite{hill83,bs95},
networks of monopoles and strings \cite{martin},
necklaces \cite{berevile97} and vortons \cite{masperi}.
Only strings and  necklaces will be considered here, while a
more extended discussion can be found in more detailed reviews 
\cite{bs00,bbv98}.

\underline{{\it Ordinary strings}}

Strings can generate UHECRs with energy less than $m_X\sim \eta$ (the
scale of symmetry breaking) if there are configurations in which microscopic 
or macroscopic portions of strings annihilate. It was shown 
\cite{shellard87,gillkibble94} 
that self-intersection events provide a flux of UHECRs which is much smaller 
than required. The same conclusion holds for intercommutation between strings. 

The efficiency of the process can be enhanced by multiple loop fragmentation: 
as a nonintersecting closed loop oscillates and radiates its energy away, the
loop configuration gradually changes. After the loop has lost a substantial 
part of its energy, it becomes likely to self-intersect and fragment into
smaller and smaller loops, until the typical size of a loop 
becomes comparable with the string width $\eta$ and the energy 
is radiated in the form of X-particles. Although the process of loop 
fragmentation is not well understood, some analytical approximations 
\cite{bbv98} show that appreciable UHECR fluxes imply
utterly large gamma ray cascade fluxes (see however \cite{bs00}).

Another way of liberating X-particles is through cusp annihilation 
\cite{brand87}, but the corresponding UHECR flux is far too 
low \cite{batta89,gillkibble94} compared with observations.

The idea that long strings lose energy mainly through 
formation of closed loops was recently challenged in the simulations of Ref.
\cite{vincent}, which show that the string can 
produce X-particles directly and that this process dominates over the
generation of closed loops. This new picture was recently questioned in Refs.
\cite{ms98,olum00}.

Even if the results of Ref. \cite{vincent} are correct however, they 
cannot solve the problem of UHECRs \cite{bbv98}: in fact the typical 
separation between two segments of a long string is comparable with the 
Hubble scale, so that UHECRs would be completely absorbed.
If by accident a string is close to us (within a few tens Mpc) then 
the UHECR events would appear to come from a filamentary region of 
space, implying a large anisotropy which is not observed. 
Even if the UHECR particles do not reach us, the gamma
ray cascade due to absorption of UHE gamma rays produced at large distances 
imposes limits on the efficiency of direct production of X-particles by
strings.

\underline{{\it Necklaces}}

Necklaces are formed when the following symmetry breaking pattern is
realized: $G\to H\times U(1) \to H\times Z_2$. In this case each monopole 
gets attached to two strings (necklace)\cite{berevile97}.
\begin{figure}[t]
  \vspace{7.0cm}
  \includegraphics{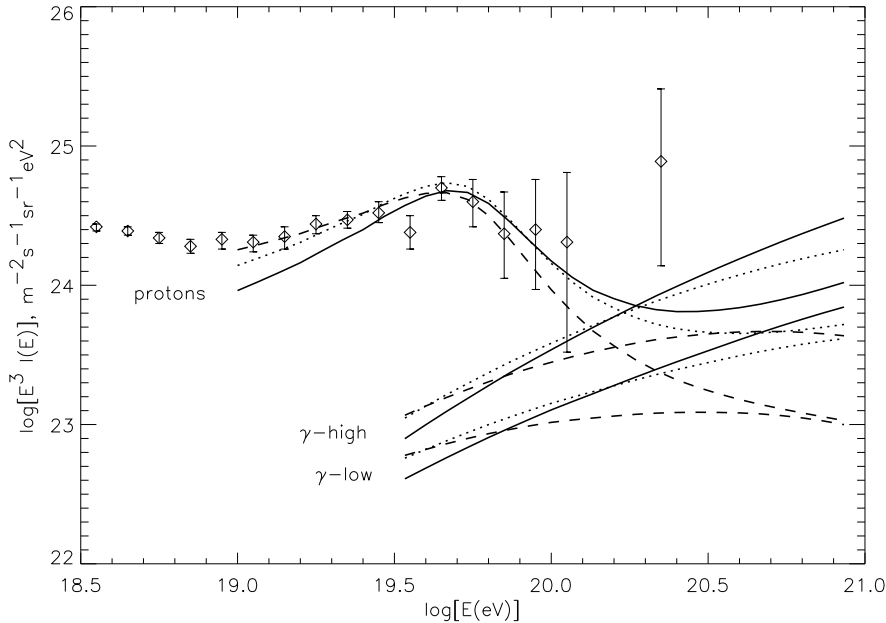}
  \caption{\it Fluxes of UHECRs from necklaces \cite{bbv98}.
    \label{fig5} }
\end{figure}
The critical parameter that defines the dynamics of this network is the 
ratio $r=m/\mu d$ where $m$ is the monopole mass and $d$ is the typical 
separation between monopoles (e.g. the length of a string segment).
If the system evolves toward a state where $r\gg 1$,
the distance between the monopoles decreases and in the 
end the monopoles annihilate, with the production of X-particles and
their decay to UHECRs. The rate of generation of X-particles is easily found
to be ${\dot n}_X\sim r^2\mu/t^3 m_X$. The quantity $r^2\mu$ is upper 
limited by the cascade radiation, given by $\omega_{cas}=\frac{1}{2} 
f_\pi r^2 \mu =\frac{3}{4} f_\pi r^2\mu/t_0^2$ ($f_\pi\sim 0.5-1$).
The typical distance from the Earth at which the monopole-antimonopole 
annihilations occur is comparable with the typical separation between 
necklaces, $D\sim\left(\frac{3f_\pi\mu}{4t_0^2\omega_{cas}}\right)^{1/4} > 10 
(\mu/10^6 GeV^2)^{1/4}$ kpc.

Clearly, necklaces provide an example in which the typical separation between 
defects is smaller than the pathlength of gamma rays and protons at 
ultra-high energies.
Hence necklaces behave like a homogeneous distribution of sources, so that
the proton component has the usual GZK cutoff. This 
component dominates the UHECR flux up to $\sim 10^{20}$ eV, while 
at higher energies gamma rays take over. The fluxes obtained 
in Ref. \cite{bbv98} are reported in Fig. 4, where the
SUSY-QCD fragmentation functions \cite{berekac98} were used. 
The dashed lines are for $m_X=10^{14}$ GeV, the dotted lines for $m_X=
10^{15}$ GeV and the solid lines for $m_X=10^{16}$ GeV.
The two curves for gamma rays refer to two different assumptions about the 
radio background at low frequencies \cite{protradio}.

\subsubsection{Cosmological relic particles}

Super heavy particles with very long lifetime can be produced in the early 
universe and generate UHECRs at present 
\cite{berekacvil97,kr98,ckr98,kt98,ktrev}. In order to keep the same 
symbolism used in previous sections, we will call these particles X-particles.

The simplest mechanism of production of X-particles in the 
early universe is the {\it gravitational production} \cite{zelsta72}:
particles are produced naturally in a time variable gravitational field
or indeed in a generic time variable classical field. In the gravitational
case no additional coupling is required (all particles interact 
gravitationally). If the time variable field is the inflaton field $\phi$, 
a direct coupling of the X-particles to $\phi$ is needed. 
In the gravitational production inflation is 
not required a priori, and indeed it reduces the effect. It can be shown 
that at time t, gravitational production can only generate X-particles 
with mass $m_X\leq H(t)\leq  m_\phi$, where $H(t)$ is the Hubble constant
and $m_\phi$ is the inflaton mass. The authors in Ref. \cite{ckr98} and
\cite{kt98} demostrated that the fraction 
of the critical mass contributed by X-particles with $m_X\sim 10^{13}$ GeV 
produced gravitationally may be $\Omega_X\sim 1$. 

If the X-particles are directly coupled to the inflaton field, they can be 
effectively generated during preheating \cite{kls94,felder98}.
Alternative mechanisms for the production of 
X-particles are based on non-equilibrium thermal generation during the 
preheating stage \cite{berekacvil97}.

As mentioned in the beginning of this section, in order for X-particles 
to be useful dark matter candidates and generate UHECRs they need to be long 
lived (for a possible annihilation scenario see Ref. \cite{dick}). 
The gravitational coupling by itself induces a lifetime much shorter 
than the age of the universe for the range of masses which we are interested 
in. Therefore, in order to have long lifetimes, additional symmetries must be 
postulated: for instance discrete gauge symmetries can protect X-particles 
from decay, while being very weakly broken, perhaps by instanton effects 
\cite{kr98}. These effects can allow decay times larger than 
the age of the universe, as shown in \cite{hama98}. The slow decay of 
X-particles produces UHECRs.
The interesting feature of this model is 
that X-particles cluster in the galactic halo, as cold dark matter 
\cite{bbv98}. 
Hence UHECRs are expected to be produced locally, with no absorption, and as 
a consequence the observed spectra are nearly identical to the emission 
spectra, and therefore dominated by gamma rays. The very flat 
spectra and the gamma ray composition are two of the signatures.
The calculations of the expected fluxes have been performed in 
\cite{bbv98,bsarkar98,blasi99,sarkar}. 
The strongest signature of the model is a slight anisotropy due to the 
asymmetric position of the sun in the Galaxy \cite{dubo98,bbv98,sarkar1}.
More recently a detailed evaluation of the amplitude and phase of the first 
harmonic has been carried out in \cite{beremik99} and \cite{medwat99}.
The two papers agree that the present data is still consistent with the 
anisotropy expected in the model of X-particles in the halo. 
Small scale anisotropies do not find an easy explanation in TD models,
with possibly the exception of the SH relics \cite{blsh00}.

\section{Hints of New Physics?}\label{sec:liv}

The possibility that at sufficiently high energies some deviations from
known Physics may occur is of particular interest for UHECRs. Some attempts
have been made to explain the events above $10^{20}$ eV as a
manifestation of some kind of Physics beyond the Standard Model of particle
interactions. 
These suggestions became even more interesting after the recent claims for
correlations of the arrival directions of UHECRs with objects at large 
redshift. Some of these correlations are still subject of debate 
\cite{correla}. More recent studies result in a 
quite intriguing correlation with BL Lacs \cite{tkachev}. Two of the 
BL Lacs in the sample used in \cite{tkachev} are in the error box of the
two triplets of events detected by AGASA, and correspond to a distance
of $\sim 600$ Mpc, much larger than the pathlength for photopion production
of protons (however more than half of the objects used in Ref. \cite{tkachev} 
have unknown redshift [Tkachev, private communication]).

The absence of bright nearby counterparts to the UHECR events has first 
inspired theoretical proposals of neutrinos as primary particles, since these
particles are not affected by the presence of the CMB and can therefore 
propagate on cosmological distances without energy losses other than those
due to the cosmic expansion. However, the small cross section of neutrinos 
makes them unlikely primaries: they simply fail to generate the observed
showers in the atmosphere. There is one caveat in this argument: the neutrino
nucleon cross section at center of mass energies above the electroweak (EW) 
scale has not been measured, so that the argument above is based on the 
extrapolation of the known cross sections, and simply limited by the weak 
unitarity bound \cite{weiler}. 
It has been proposed \cite{domokos} that an increase in the number of
degrees of freedom above the EW scale would imply the increase of the 
neutrino nucleon cross section above the standard model prediction. In 
particular, in the theories that predict unification of forces at 
$\sim 1$ TeV scale with large extra dimensions, introduced to solve the
hierarchy problem \cite{arkani}, these additional degrees of freedom arise 
naturally, and imply a neutrino-nucleon cross section that increases linearly
with energy, reaching hadronic levels for neutrino energies of $\sim 10^{20}$ 
eV (for string scale at 1 TeV). The calculations in Ref. \cite{plu} show
that the cross section remains too small to explain vertical showers in
the atmosphere (for possible upper limits on this cross section see Ref.
\cite{tyler}).

An alternative, even more radical proposal to avoid the prediction of a GZK 
cutoff in the flux of UHECRs consists in postulating a tiny violation of
Lorentz invariance (LI). The main effect of this violation is that some 
processes may become kinematically forbidden \cite{livio}.
In particular, photon-photon pair production and photopion production
may be affected by LI violation. The absence of the GZK cutoff would then 
result from the fact that the threshold for photopion production disappears 
and the process becomes kinematically not allowed.
It is suggestive that the possibility of a small violation
of LI has also been proposed to explain the apparent absence of an absorption
cutoff in the TeV gamma ray emission from Markarian-like objects
\cite{protheroe} (see \cite{bererecent} for a critical view of this
possibility). 
  
\section{Conclusions}\label{sec:conclusion}

The increasing evidence for a flux of UHECRs exceeding the theoretical 
expectations for extragalactic sources has fueled interest in several 
models. These models aim to satisfy all the requirements imposed by 
observations on fluxes, spatial anisotropies and composition. At present, 
however there is no obvious successful model. The situation 
might change with the availability, soon to come, of quite larger statistics 
of events, that will be achievable by experiments like the Pierre Auger 
Project \cite{cronin} and EUSO/Airwatch/OWL \cite{scarsi,nasa}. 

These future experimental efforts will be crucial mainly in three respects:
1) the increase of statistics by a factor 100 for Auger and even more for
the space based experiments will allow to strongly constrain theoretical
models, and check whether the present excess is a $(2-3)\sigma$ fluctuation or
a physical effect. Moreover the small scale anisotropies, if real, will 
become stronger and a correlation function approach will definitely become
appropriate to the analysis of the events; 2) the full sky coverage will 
finally allow a test of models based on local extragalactic sources and on 
galactic sources of UHECRs, through the measurement of the large scale 
anisotropy, that is currently spoiled by the limited spatial exposure; 3) a
better determination of the composition of the UHECR events will represent a 
smoking gun either in favour of or against models: TD models would be ruled
out if no gamma rays are found or if heavy nuclei represent the main 
component. On the other hand, iron-dominated composition would point toward
a possible galactic origin, possibly related to neutron stars.

The confirmation, on statistical grounds, of the association of UHECRs
to distant cosmological objects like BL Lacs would represent a very strong
indication of physics beyond the standard model. Either new interactions 
or some modification of fundamental physics would be needed to explain 
such a result.


\begin{thebibliography}{99}

\bibitem{greisenzk66} K. Greisen, Phys. Rev. Lett. {\bf 16}, 748 (1966);
G.T. Zatsepin and V.A. Kuzmin, Sov. Phys. JETP Lett. {\bf 4}, 78 (1966).

\bibitem{linsley} J. Linsley, Phys. Rev. Lett. {\bf 10}, 146 (1963).

\bibitem{watson91} 
M.A. Lawrence, R.J.O. Reid and A.A. Watson, J. Phys. G. Nucl.
Part. Phys. {\bf 17}, 773 (1991).

\bibitem{efimov91} 
N.N. Efimov {\it et al}: Ref. Proc. International Symposium on 
{\it Astrophysical Aspects of the most energetic cosmic rays}, eds M. Nagano
and F. Takahara (World Scientific, Singapore), p. 20 (1991).

\bibitem{agasa01}
M. Teshima, proceedings of TAUP 2001, Laboratori Nazionali del
Gran Sasso, L'Aquila, Sept. 8-12, 2001.

\bibitem{tak99} M. Takeda {\it et al} Astrophys. J. {\bf 522}, 225 (1999).

\bibitem{tak98} M. Takeda {\it et al}, Phys. Rev. Lett. {\bf 81}, 1163 (1998).

\bibitem{ha94} N. Hayashida {\it et al}, Phys. Rev. Lett. {\bf 73}, 3491 
(1994).

\bibitem{bird93} D.J. Bird {\it et al}, Phys. Rev. Lett. {\bf 71}, 3401 
(1993). 

\bibitem{bird94}  D.J. Bird {\it et al},
Astrophys. J. {\bf 424}, 491 (1994).

\bibitem{bird95} 
D.J. Bird {\it et al}, Astrophys. J. {\bf 441}, 144 (1995). 

\bibitem{kieda99} 
D. Kieda {\it et al}, {\it HiRes Collaboration} 1999 Proc. of 26th ICRC,
Salt Lake City, Utah.

\bibitem{uchihori00} 
Y. Uchihori, M. Nagano, M. Takeda, M. Teshima, J. Lloyd-Evans and
A.A. Watson,  Astropart. Phys. {\bf 13}, 151 (2000).

\bibitem{stanev}
T. Stanev, proceedings of the Vulcano Workshop ``Frontier
Objects in Astrophysics and Particle Physics'', Vulcano, May 21-27, 2000.

\bibitem{halzen} 
F. Halzen, R. Vazques, T. Stanev and H.S. Vankov, Astropart. 
Phys. {\bf 3}, 151 (1995).

\bibitem{yoda98} S. Yoshida and H. Dai, J. Phys. G {\bf 24}, 905 (1998).

\bibitem{zas2000}
M. Ave, J.A. Hinton, R.A. Vazquez, A.A. Watson, E. Zas 
Phys. Rev. Lett. {\bf 85}, 2244 (2000).

\bibitem{hiresICRC}
Numerous contributions have been presented by the HiRes Collaboration at the
ICRC2001.

\bibitem{bereICRC}
V. Berezinsky, A.Z. Gazizov, S.I. Grigorieva, preprint hep-ph/0107306.

\bibitem{blanton}
M. Blanton, P. Blasi and A.V. Olinto, Astrop. Phys. {\bf 15}, 275 (2001).

\bibitem{grigo88}
V.S. Berezinsky and S.I. Grigorieva, A\&A, {\bf 199}, 1 (1988).

\bibitem{stanev00}
T. Stanev {\it et al}, Phys. Rev. {\bf D62}, 093005 (2000).

\bibitem{pscz}
W. Saunders {\it et al}, preprint astro-ph/0001117.

\bibitem{cfa2}
J.P. Huchra, M.J. Geller and H.J. Corwin Jr., Astroph. J. {\bf 70} 687 (1995).

\bibitem{bac_wax}
J.N. Bachall and E. Waxman, Astroph. J. {\bf 542}, 542 (2000).

\bibitem{hayashida99}
N. Hayashida, et al., Appendix to Astroph. J. {\bf 522}, 225 (1999). 

\bibitem{olinto} 
A.V. Olinto, Phys. Rep. {\bf 333}, 329 (2000).

\bibitem{blandford}
R.D. Blandford, Phys. Scripta {\bf T85}, 191 (2000).

\bibitem{hillas}
A.M. Hillas, Ann. Rev. Astron. Astrop. {\bf 22}, 425 (1984).

\bibitem{GRB}
E. Waxman, Phys. Rev. Lett. {\bf 75}, 386 (1995); ibid., Astroph. J. 
{\bf 452}, L1 (1995); M. Vietri, Astroph. J. {\bf 453}, 883 (1995).

\bibitem{dar}
A. Dar, preprint astro-ph/9901005; F.W. Stecker, Astropart. Phys. 
{\bf 14}, 207 (2000).

\bibitem{bible} Berezinksy, V.S., Bulanov, S.V., Dogiel, V.A., Ginzburg, V.L.,
Ptuskin, V.S., {\it Astrophysics of Cosmic Rays} (Amsterdam: North
Holland, 1990).

\bibitem{venkatesan}
A. Venkatesan, M.C. Miller and A.V. Olinto, Astroph. J. {\bf 484}, 323 (1997).

\bibitem{boe}
P. Blasi, R.I. Epstein and A.V. Olinto, 
Astrophys. J. {\bf 533}, L123 (2000).

\bibitem{lazarian}
E.M. de Gouveia Dal Pino and A. Lazarian,
Astrophys. J. {\bf 536}, L31 (2000)

\bibitem{begelman}
M.C. Begelman, Astroph. J. {\bf 493}, 291 (1998). 

\bibitem{bedn}
J.H. Beall and W. Bednarek, preprint astro-ph/0108447.

\bibitem{aniso}
J. Alvarez-Muniz, R. Engel and T. Stanev, in the proceedings of
ICRC2001, OG1.03, Hamburg, Germany, 07-15 August 2001; 
S. O'Neill, A.V. Olinto and P. Blasi, in the proceedings of
ICRC2001, OG1.03, Hamburg, Germany, 07-15 August 2001 (preprint 
astro-ph/0108401).

\bibitem{thorne}
K.S. Thorne, R.M. Price, and D. MacDonalds, 1986 Black Holes: The Membrane
Paradigm (New Haven: Yale Press) and references therein.

\bibitem{magorrian}
J. Magorrian {\it et al}, Astron. J. {\bf 115}, 2528 (1998).

\bibitem{boldt}
E. Boldt and P. Ghosh, MNRAS {\bf 307}, 491 (1999).

\bibitem{levinson}
A. Levinson, preprint hep-ph/0002020.

\bibitem{biermann}
J.P. Rachen and P.L. Biermann, A\& A {\bf 272}, 161 (1993).

\bibitem{slb} G. Sigl, M. Lemoine and P.L. Biermann,  Astropart. Phys. 
{\bf 10}, 141 (1999).

\bibitem{blasiolinto99} 
P. Blasi and A.V. Olinto, Phys. Rev. {\bf D59}, 023001 (1999).

\bibitem{ahn} 
E.J. Ahn, P.L. Biermann, G. Medina-Tanco and T. Stanev, preprint
astro-ph/9911123.

\bibitem{bs00} P. Bhattacharjee and G. Sigl, Phys. Rep. {\bf 327}, 109 (2000).

\bibitem{vilshe94} 
A. Vilenkin and E.P.S. Shellard, {\it Cosmic Strings and Other
Topological Defects}, Cambridge University Press, Cambridge (1994).

\bibitem{hsw87} 
C.T. Hill, D.N. Schramm and T.P. Walker, Phys. Rev. {\bf D36}, 1007 (1987).

\bibitem{rana90} 
P. Bhattacharjee and N.C. Rana, Phys. Lett. {\bf B246}, 365 (1990).

\bibitem{hill83} C.T. Hill, Nucl. Phys. {\bf B224}, 469 (1983).

\bibitem{bs95} P. Bhattacharjee and G. Sigl, Phys. Rev. {\bf D51}, 4079 (1995).

\bibitem{martin} 
V.S. Berezinsky, X. Martin and A. Vilenkin, Phys. Rev. {\bf D56}, 2024 (1997).

\bibitem{berevile97} 
V.S. Berezinsky and A. Vilenkin,  Phys. Rev. Lett. {\bf 79}, 5202 (1997).

\bibitem{masperi} 
L. Masperi and B. Silva, Astropart. Phys. {\bf 8}, 173 (1998).

\bibitem{bbv98} V.S. Berezinsky, P. Blasi and A. Vilenkin, Phys. Rev. 
{\bf D58}, 103515 (1998).

\bibitem{shellard87} E.P.S. Shellard, Nucl. Phys. {\bf B283}, 624 (1987).

\bibitem{gillkibble94} 
A.J. Gill and T.W.B. Kibble, Phys. Rev. {\bf D50}, 3660 (1994).

\bibitem{brand87} R. Brandenberger, Nucl. Phys. {\bf B293}, 812 (1987).

\bibitem{batta89} P. Bhattacharjee, Phys. Rev. {\bf D40}, 3968 (1989).

\bibitem{vincent} 
G. Vincent, N. Antunes and M. Hindmarsh, Phys. Rev. Lett. 
{\bf 80}, 2277 (1998).

\bibitem{ms98} J.N. Moore and E.P.S. Shellard, preprint astro-ph/9808336.

\bibitem{olum00} K. Olum and J.J. Blanco-Pillado, Phys. Rev. Lett. {\bf 84}, 
4288 (2000).

\bibitem{berekac98} V.S. Berezinsky and M. Kachelriess, Phys. Lett. 
{\bf B434}, 61 (1998).

\bibitem{protradio} 
R.J. Protheroe and P.L. Biermann, Astropart. Phys. {\bf 6}, 45 (1996).

\bibitem{berekacvil97} 
V.S. Berezinsky, M. Kachelriess and A. Vilenkin, Phys. Rev. Lett. 
{\bf 79}, 4302 (1997).

\bibitem{dick}
P. Blasi, R. Dick and E.W. Kolb, preprint astro-ph/0105232.

\bibitem{kr98} V.A. Kuzmin and V.A. Rubakov, Yadern. Fiz. {\bf 61},  
1122 (1998).

\bibitem{ckr98} 
D.J.H. Chung, E.W. Kolb and A. Riotto, Phys. Rev. {\bf D59}, 023501 (1998).

\bibitem{kt98}  V.A. Kuzmin and I.I.Tkachev, JETP Lett. {\bf 69}, 271 (1998).

\bibitem{ktrev}  V.A. Kuzmin and I.I.Tkachev, Phys. Rep. {\bf 320}, 199 (1999).

\bibitem{zelsta72} 
Ya.B. Zeldovich and A.A Starobinsky, Soviet Phys. JETP {\bf 34}, 1159 (1972). 

\bibitem{kls94} 
L. Kofman, A. Linde and A. Starobinsky, Phys. Rev. Lett. {\bf 73}, 
3195 (1994).

\bibitem{felder98} 
G. Felder, L. Kofman and A. Linde, 1998 preprint astro-ph/9812289.

\bibitem{hama98} 
K. Hamaguchi, Y. Nomura and T. Yanagida, Phys. Rev. {\bf D58}, 103503 (1998). 

\bibitem{bsarkar98} 
M. Birkel and S. Sarkar, Astropart. Phys. {\bf 9}, 297 (1998). 

\bibitem{blasi99} P. Blasi, Phys. Rev. {\bf D60}, 023514 (1999).

\bibitem{sarkar}
S. Sarkar, `COSMO-99, Third Intern. Workshop on Particle Physics and 
the Early Universe' pages 77-91, Trieste, 27 Sep-3 Oct 1999.

\bibitem{dubo98} 
S.L. Dubovsky and P.G. Tynyakov, Pis'ma Zh. Eksp. Teor. Fiz. 
{\bf 68}, 99 [JETP Lett. {\bf 68}, 107] (1998).

\bibitem{sarkar1}
W. Evans, F. Ferrer and S. Sarkar, preprint astro-ph/0103085.

\bibitem{beremik99} 
V.S. Berezinsky and A. Mikhailov, Phys. Lett. {\bf B449}, 237 (1999).

\bibitem{medwat99} 
G.A. Medina-Tanco and A.A. Watson, Astropart. Phys. {\bf 12}, 25 (1999).

\bibitem{blsh00} 
P. Blasi and R.K. Sheth, R.K., Phys. Lett. {\bf B486}, 233 (2000).

\bibitem{correla}
G.R. Farrar and P.L. Biermann, Phys. Rev. Lett. {\bf 81}, 3579 (1998);
C.M. Hoffman, Phys. Rev. Lett. {\bf 83}, 2471 (1999);
G.R. Farrar and P.L. Biermann, Phys. Rev. Lett. {\bf 83}, 2472 (1999);
G. Sigl {\it et al}, preprint astro-ph/0008363;
A. Virmani {\it et al}, preprint astro-ph/0010235.

\bibitem{tkachev}
P.G. Tinyakov and I. I. Tkachev, preprint astro-ph/0102476

\bibitem{weiler}
H. Goldberg and T. Weiler, Phys. Rev. {\bf D59}, 113005 (1999).

\bibitem{domokos}
G. Domokos and S. Kovesi-Domokos, Phys. Rev. Lett. {\bf 82}, 1366 (1999). 

\bibitem{tyler}
C. Tyler, A.V. Olinto and G. Sigl, Phys. Rev. {\bf D63}, 055001 (2001).

\bibitem{arkani}
N. Arkani-Hamed, S. Dimopoulos and G. Dvali, Phys. Rev. {\bf D59}, 086004
(1999).

\bibitem{plu}
M. Kachelriess and M. Plumacher, Phys. Rev. {\bf D62}, 103006 (2000).

\bibitem{livio}
S. Coleman and S.L. Glashow, Nucl. Phys. {\bf B574}, 130 (2000);
L. Gonzales-Mestres, Nucl. Phys. B (Proc. Suppl.) {\bf 48}, 131 (1996);
R. Aloisio, P. Blasi, P. Ghia and A.F. Grillo, Phys. Rev. {\bf D62}, 053010
(2000).

\bibitem{protheroe}
R.J. Protheroe, H. Meyer, Phys. Lett. {\bf B493}, 1 (2000).

\bibitem{bererecent}
V.S. Berezinsky, preprint astro-ph/0107306.

\bibitem{cronin} 
J.W. Cronin, Nucl. Phys. B. (Proc. Suppl.) {\bf 28B}, 213 (1992).

\bibitem{scarsi}
See web page: http://www.ifcai.pa.cnr.it/~EUSO/home.html

\bibitem{nasa} 
R.E. Streitmatter, Proc. of {\it Workshop on Observing Giant 
Cosmic Air Showers from $>10^{20}$ eV Particles from Space}, eds. 
Krizmanic, J.F., Ormes, J.F., and Streitmatter, R.E. (AIP Conference
Proceedings 433, 1997).

\end{thebibliography}
\end{document}